\documentclass[aps,superscriptaddress,showpacs,nofootinbib,floatfix]{revtex4}
\usepackage{graphicx}
\usepackage{bm,color,ulem, soul}
\usepackage{amssymb}
\newcommand{\me}{\varepsilon}

\newcommand{\dd}{\mathrm{d}}

\newcommand{\QGP}{\textrm{\scriptsize QGP}}
\newcommand{\HG}{\textrm{\scriptsize HG}}

\newcommand{\bc}[1]{[{\bf \color{blue}{baoyichen}}]}

\begin{document}
\author{Jingjing Wang}
\author{Baoyi Chen}
\author{Yunpeng Liu}
\email{Email:yunpeng.liu@tju.edu.cn}
\affiliation{Department of Applied Physics, Tianjin University, Tianjin 300350, China}
\pacs{25.75.Nq, 12.38.Mh, 25.75.-q}
\title{Finite size effect on gluon dissociation of $J/\psi$ in relativistic heavy ion collisions}
\begin{abstract}
	Thermal quantities, including the the entropy density and gluon spectrum, of quark matter within a box that is finite in the longitudinal direction are calculated using a bag model. Under the assumption of entropy conservation, the corresponding gluon dissociation rate of $J/\psi$ is studied. It reaches a maximum at a certain longitudinal size $L_m$, below which the suppression is weak even if the temperature becomes higher than that without the finite size effect, and above which the dissociation rate approaches to the thermodynamic limit gradually with increasing longitudinal size of the fireball.
\end{abstract}
\maketitle

\section{Introduction}
The quark-gluon plasma (QGP) is widely accepted as the state of matter with strong interaction at high temperature and/or high density in theory~\cite{Hagedorn:346206, Karsch:2000kv, Ding:2015ona, Pisarski:1983ms, Fukushima:2010bq, Stephanov:1998dy, Kharzeev:2007tn}, and it is also studied for decades in experiments of relativistic heavy ion collisions at a typical space scale of $10$ fm and a typical time scale of $10$ fm$/c$~\cite{Gonin:1996wn,Luo:2017faz, Adcox:2001jp, Adare:2006ti,Chatrchyan:2011sx}. Many probes are suggested to detect the new state of matter in such small scales, one of which is the $J/\psi$ suppression~\cite{Matsui:1986dk, Andronic:2009qf}. Different from light hadrons, part of $J/\psi$s survive the hot medium due to its large binding energy, and its suppression carries the information of the fireball at its early stage~\cite{Karsch:2005nk, Xu:1995eb, Brambilla:1999xf, Escobedo:2013tca, CS:2018mag}. In a microscopic view, the leading order process of $J/\psi$ suppression is the gluon dissociation through a dipole interaction~\cite{Peskin:1979va, Grandchamp:2001pf, Zhu:2004nw}. Later on, the inverse process is also suggested to play an important role in nucleus-nucleus collisions when charm pairs are abundant~\cite{Thews:2000rj,Grandchamp:2001pf,Yan:2006ve}. High order processes were also considered~\cite{Chen:2018dqg}. The study also extends from $J/\psi$\ to its excited states~\cite{Wong:1996wm, Andronic:2009qf, Grandchamp:2001pf} and their analogs $\Upsilon$s~\cite{Lansberg:2006dh}, and even $B_c$ mesons~\cite{Schroedter:2000ek, Liu:2012tn, Irfan:2015qxa, CMS:2022sxl,Zhao:2022auq, Wu:2023djn}.

Most theories focus on the thermodynamic quantities in the thermodynamic limit, while the size of the fireball is small, especially in the longitudinal direction at early stage of the fireball. The finite size effect is attracting more attention in relativistic heavy ion collisions~\cite{Xu:2023epi, Deb:2020qmx, Yamamoto:2009ey, Pal:2023aai}. This may be less important for light hadrons, which are produced in later stage of the fireball. However, for particles like $J/\psi$ that carries the information of the fireball at early stage, it is necessary to discuss the finite size effects, which come from the fact that the wave function of particles is within a finite size. In Ref.~\cite{Guo}, the gluon dissociation of $J/\psi$ with finite size effect is compared with that at infinity space at exactly the same temperature. It is found that even if the temperature of the medium is high, the suppression could still be  small due to a small size in space. It is also found that the binding energy of a $J/\psi$ becomes smaller when its wave function is constrained in a sphere rigoriously~\cite{Cheng:2019xlx}.

Heavy quarkonia are affected by the finite size effects via two aspects.
\begin{enumerate}
	\item Direct effects. The finite size of the fireball changes the gluon distribution, which directly modifies the decay rates of $J/\psi$ at a certain temperature.
	\item Indirect effects.  The finite size of the fireball changes the thermal properties of the medium,  such as the equation of state and consequently the critical temperature of deconfinement, which affect the total suppression of $J/\psi$s in heavy ion collision.
\end{enumerate}
The finite size effect on gluon distribution and the corresponding $J/\psi$ dissociation rate have been studied in previous work~\cite{Guo, Cheng:2019xlx}. This work focus on the finite size effect on the bulk medium, especially its equation of state and the $J/\psi$ dissociation in this hot medium.

In this paper, we assume that the system is uniform and infinitely large in the transverse direction, but with a finite size $L=2ct$ in the longitudinal direction proportional to the time $t$ after collision, because all wave functions of partons vanishes on the boundary of the light cone of the nucleus collision. It is also assumed that the expansion is adiabatic, which indicates conserved entropy, as is usually assumed in a 2+1 D hydrodynamic models. As a result, the entropy density $s$ evolves as $s\propto 1/t$. Therefore a comparison of quantities between the cases with and without the finite size effect at the same time $t$ is equivalent to a comparison at the same entropy density $s$. For this reason, we discuss the finite size effect on the thermal quantities, especially on the entropy density $s$, in section II. The dissociation rate of $J/\psi$ with the finite size effects is calculated and discussed in section III. In Section IV, we shortly summarize the results. For simplicity, we only consider the gluon dissociation process of $J/\psi$s at middle rapidity as in Ref.~\cite{Guo}, and the natural units $\hbar=c=k_{\mathrm{B}}=1$ are adopted.

\section{Entropy density with a finite longitudinal size}\label{section2}
The medium is described by a bag model with a first-order phase transition.~\cite{Han} The grand thermal potential reads
\begin{eqnarray*}
    J(T, L)=\left\{\begin{array}{ll}
-T \ln {Z}_{\QGP}^{0}+BV, & \text {for QGP } \\
-T \ln {Z}_{\HG},& \text {for hardon gas}
\end{array}\right.
\end{eqnarray*}
where $Z_{\QGP}^0$ and $Z_{\HG}$ are the partition functions of ideal parton gas and that of ideal hadron gas. Partons includes massless gluons and u, d quarks, and massive $s$ quarks with its mass $m_s=150$ MeV. Hadrons are those in the particle list with its mass below $2$ GeV. The temperature, volume and longitudinal size of the fireball are denoted as $T$, $V$, and $L$, respectively, and $B$ is the bag constant, which is taken as $B=(236\textrm{ MeV})^{4}$ in our numerical calculations leading to a critical temperature $T_c=165$ MeV in an infinite space.

For each particle $i$ with mass $m_i$ in the ideal gas, the corresponding partition function $Z_i$ satisfies
\begin{eqnarray}
   \ln {Z}_{i} &=&\pm \int_{m_i}^{+\infty} D(\varepsilon_{i}) \ln \left(1 \pm e^{-\beta \varepsilon_i}\right) \mathrm{d} \varepsilon_{i}.\label{eq_lnZi_1}
\end{eqnarray}
The upper and lower signs are for fermions and bosons, respectively, here and below. The density of states at particle energy $\me_i$ is~\cite{Han}
\begin{eqnarray}
   D(\me_i)&=& \frac{g_iV\me_i}{2\pi L}\left[\frac{p_i L}{\pi}\right],
\end{eqnarray}
with $g_i$ counting the inner degree of freedom of particle $i$, and  $p_i=\sqrt{\me_i^2-m_i^2}$ being the magnitude of its momentum. The square brackets here and below stand for the least integer function, which comes from the summation on the discrete longitudinal momentum $p_z=\frac{n_z\pi}{L}$ with $n_z=1,2,3\cdots$. By substituting the density of states into Eq.~(\ref{eq_lnZi_1}), we have
\begin{eqnarray}
   \ln Z_i &=&\pm \frac{g_{i} V}{2 \pi\beta^{3}\Lambda} \int_0^{+\infty}  x\left[\frac{x \Lambda}{\pi}\right] \ln \left(1 \pm e^{-\xi_i}\right)\mathrm{d} x, 
\end{eqnarray}
with $\xi_i=\sqrt{x^2+\bar{m}_i^2}$, $\bar{m}_i=\beta m_i$, and $\beta=1/T$. 

Accordingly, the entropy of particle $i$ is
\begin{eqnarray*}
   S_i&=& \frac{g_iV}{2\pi\beta^3\Lambda}\int_0^{+\infty}x\left[\frac{x\Lambda}{\pi}\right]\left(\pm\ln\left(1\pm e^{-\xi_i}\right)+\frac{\xi_i}{e^{\xi_i}\pm 1}\right)\mathrm{d} x.
\end{eqnarray*}
Taking the limit $L\rightarrow +\infty$ (technically just by removing the square brackets) in all the equations above, the results in infinite space will be recovered. In the other limit $L\rightarrow 0$, 
the constraint in space increases the energy of the ground state. As a result, the differences between bosons and fermions are negligible, and thus it can be approximately described in a classical limit.
For a massless classical particle, its entropy is 
\begin{eqnarray}
   S_i^{\textrm{\scriptsize cl}}&=& \frac{g_iV}{2\pi\beta^3}\frac{1}{\pi}\frac{\eta q}{1-q}\left(\eta^2\frac{1+q}{(1-q)^2}+\frac{3\eta}{1-q}+3\right),
\end{eqnarray}
with $\eta=\pi/\Lambda$, $q=e^{-\eta}$, and $\Lambda=LT$. This is a good approximation of the entropy of a massless quantum particle in the sense that the error is within 10\% in the whole range of $L$. 
In the small $L$ limit, that is $\Lambda\rightarrow 0$, $\eta\rightarrow +\infty$ and $q\rightarrow 0$, it gives
\begin{eqnarray}
   S_i&=& \frac{g_iA\pi}{2L^2}e^{-\frac{\pi}{\Lambda}},
\end{eqnarray}
where $A=V/L$ is the transverse area of the system. Summing up all the light partons, one obtains the total entropy per area
\begin{eqnarray}
   s_A\equiv \frac{S}{A}&\approx& \frac{G\pi}{2L^2}e^{-\frac{\pi}{\Lambda}},\label{eq_LL}
\end{eqnarray}
where $G$ is the total internal degree of freedom of light partons.
At early time of heavy ion collisions, the longitudinal size of the fireball is small, and the transverse expansion is negligible, so that the entropy per area $s_A$ keeps constant. Therefore Eq.~(\ref{eq_LL}) implies that when we go to the early limit $t\rightarrow 0$, that is $L\rightarrow 0$, we have $\Lambda\rightarrow 0$ at a speed in the order of magnitude of  $1/(\ln (1/L))$ in spite of $T\rightarrow +\infty$, with the assumption that $s_A$ is a constant. Therefore the time evolution of $T$ is that $T$ drops with $t$ at a speed slightly slower than $T\propto 1/t$ at small $L$. 

\par
\begin{figure}[!hbt]
    \centering
    \includegraphics[width=0.7\textwidth]{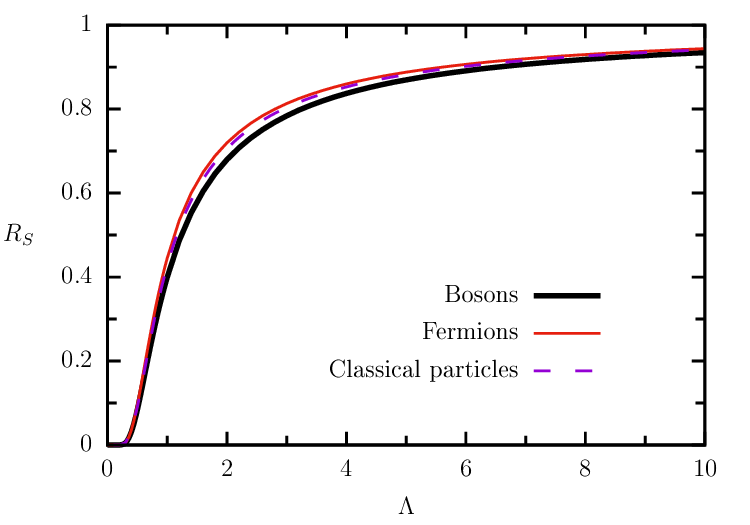}
    \caption{The ratio of entropy density of a massless particle in finite space to that in infinite space at the same temperature as a function of $\Lambda$.}
    \label{fg_sratio}
\end{figure}

The entropy of massless particles in infinite space is
\begin{eqnarray}
    S_i(\Lambda=\infty,m=0)&=&\frac{(15\mp 1)\pi^2g_iV}{360\beta^3}
\end{eqnarray}
The ratio between the entropy $S_i(\Lambda, m=0)$ of massless particles in finite space to $S_i(\Lambda=\infty, m=0)$ above only depends on $\Lambda=LT$, and is independent of $T$ itself. The ratio
\begin{eqnarray*}
   R_S(\Lambda)\equiv\frac{S_i(\Lambda,m=0)}{S_i(\Lambda=+\infty, m=0)}
\end{eqnarray*}
is shown in Fig.~(\ref{fg_sratio}). It can be seen that the ratio increases with $\Lambda$ monotonically. The asymptotic expression is
\begin{eqnarray*}
   R_S(\Lambda)&=& \left\{\begin{array}{ll}1-\frac{b_\pm}{\Lambda},&\textrm{ for large }\Lambda\\\frac{180}{(15\mp1)\pi \Lambda^3}\left(1+\frac{3\Lambda}{\pi}+\frac{3\Lambda^2}{\pi^2}\right)e^{-\frac{\pi}{\Lambda}}, & \textrm{ for small }\Lambda.\end{array}\right.
\end{eqnarray*}
with $b_\pm=\frac{7\mp1}{15\mp1}\frac{270\zeta(3)}{8\pi^3}$, that is $b_+\approx 0.561$ and $b_-\approx 0.654$, and its classical corresponding is $b_{\mathrm{cl}}=\frac{3\pi}{16}\approx 0.589$. In practice, this formula works well for $\Lambda>2$ or $\Lambda<0.5$. 
The results show that the entropy density becomes smaller when the finite size effect is included compared with that in infinite space at the same temperature, and therefore at a fixed entropy density higher temperature is expected in finite space than that in infinite space.

\begin{figure}[!hbt]
    \centering
    \includegraphics[width=0.7\textwidth]{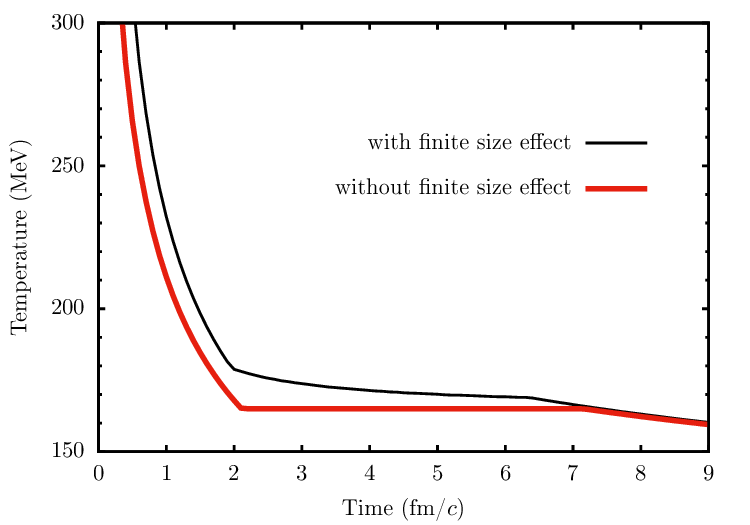}
    \caption{Time evolution of the medium temperature with and without the finite size effect. The entropy densities are taken the same in both cases.}
    \label{fg_T}
 \end{figure}
 The time evolution of the temperature $T$ is shown in Fig.~\ref{fg_T}. The $s_A$ is fixed by taking the temperature to be $T=250$ MeV at $t=0.6$ fm/c in the case without the finite size effects. In the hadrons phase with large time $t$, the finite size effect is small and negligible, and in the mixed phase the difference between the two curves shows the shift of the critical temperature~\cite{Han}. The most significant modification happens at the early time when the temperature of the medium is remarkably higher with the finite size effect, although the entropy densities are the same in two cases. 
\section{$J/\psi$ suppression}
The suppression of heavy quarkonia is attributed to gluon dissociation at the leading order, and the dissociation rate of $J/\psi$ is
\begin{eqnarray}
	&&\alpha({\bm p},  T, L)\nonumber\\
   &=& \frac{1}{EL}\sum_{n_z=1}^{+\infty}\int\frac{\dd {\bm k}_T}{(2\pi)^2\varepsilon_g}p_{\mu}k^{\mu}\sigma_{gJ/\psi }^{c\bar{c}}f_g({\bm k}_T,n_z,T,L),\nonumber
\end{eqnarray}
where $p=(E,{\bm p})$\ and $k=(\varepsilon_g,{\bm k}_T, k_z)$ are four-momenta of the $J/\psi$ and the gluon, respectively, with $k_z=\frac{\pi}{L}n_z$. The cross section of gluon dissociation process of $J/\psi$ can be calculated via the Operator Production Expansion method~\cite{Peskin:1979va, Polleri:2003kn} as
\begin{eqnarray}
	\sigma_{gJ/\psi}^{c\bar{c}}&=& A_0\frac{(\omega/\epsilon-1)^{3/2}}{(\omega/\epsilon)^5},\label{eq_8}
\end{eqnarray}
where $\omega=\frac{(p+k)^2-m^2}{2m}$\ is the energy of gluon in $J/\psi$ frame, and $m$\ is the mass of a $J/\psi$. In the above, $\epsilon=\frac{4m_c^2-m^2}{2m}$ is the threshold of the process, with the constant $A_0=2^{11}{\pi}/(27\sqrt{m_c^3\epsilon})$. The distribution function of gluons is assumed to be thermal in the lab frame
\begin{eqnarray}
	f_g({\bm k}_T, n_z, T, L)&=& \frac{g_g}{e^{\beta \varepsilon_g}-1}
\end{eqnarray}
with $g_g=16$ being the degree of freedom of spin and color, and $\epsilon_g$ is the energy of gluon in the medium frame.

\begin{figure}[!hbt]
    \centering
    \includegraphics[width=0.7\textwidth]{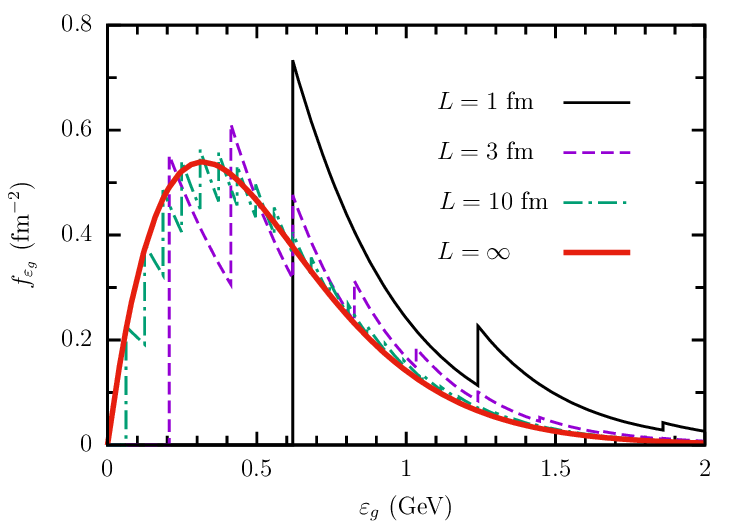}
    \caption{Finite size effect on the gluon spectrum. All curves are with the same entropy density but different longitudinal size $L$. The temperature of gluons at $L=+\infty$ is $200$ MeV.}
    \label{fg_gs}
 \end{figure}

When the system size is finite, 
the gluon spectrum is
\begin{eqnarray}
   f_{\me_g}\equiv\frac{\dd N_g}{\dd \me_g\dd V}=\frac{g_g}{2\pi L}\left[\frac{\me_g}{\me_0}\right]\frac{\me_g}{e^{\beta\me_g}-1},
\end{eqnarray}
where $N_g$ and $\me_0=\pi/L$ are the number of gluons and the energy of their ground state, respectively. 
The spectra of gluons in both finite and infinite size at the same temperature are compared in Ref.~\cite{Guo}. With the assumption that the medium expands adiabatically, a comparison at the same entropy density makes sense. For simplicity, we consider the medium as massless gluon gas without quarks firstly. This is shown in Fig.~\ref{fg_gs}. The spectrum starts to be finite at $\varepsilon_g=\varepsilon_0$, and there is a jumping point in the spectrum at every multiple of $\varepsilon_0$ because of discrete momentum values in the longitudinal direction. 
Integrating the spectrum function, one obtains the average number of gluons $N_g=\int f_{\varepsilon_g}\dd \varepsilon_g \dd V$. In a small $\Lambda$ limit, it gives $N_g=\frac{g_g A\Lambda}{2L^2}e^{-\frac{\pi}{L}}=\Lambda S_g/\pi$. As discussed in the previous section, with fixed entropy density, $\Lambda$ decreases slowly as $L$ decreases at small $L$. Therefore the number density of gluons becomes slightly smaller, when $L$ becomes smaller. Besides that, when the system becomes smaller, the spectrum shifts to the higher energy side. For gluon dissociation, the cross section is negligible when the energy of gluon is far larger than the threshold $\epsilon$. Therefore when the size is small enough, even if the temperature of the medium is high, the suppression may still be weak.

\begin{figure}[!hbt]
    \centering
    \includegraphics[width=0.7\textwidth]{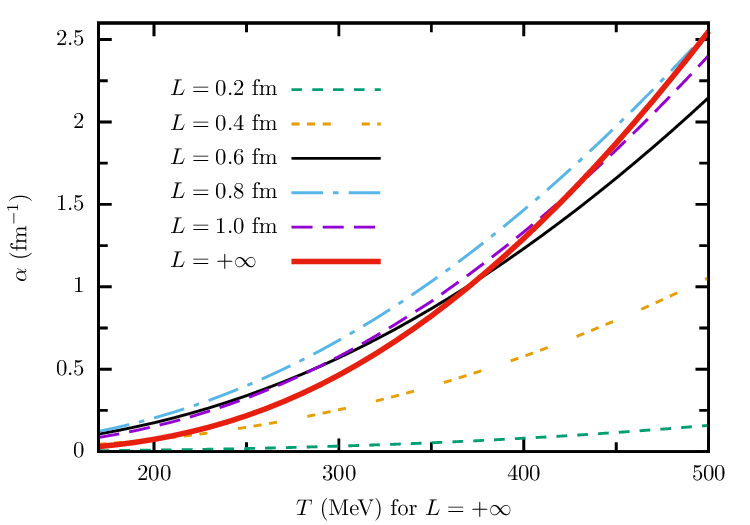}
    \caption{Comparison of the dissociation rate $\alpha$\ of $J/\psi$ with with different longitudinal size $L$ of space. Points at the same $x$-coordinate have the same entropy density. To be intuitive, we label the $x$-axis by the temperature $T$ of the infinite space instead.}
    \label{fg_alpha_L}
\end{figure}

The dissociation rate $\alpha$\ of $J/\psi$s at rest as functions of entropy density of gluon gas is shown in Fig.~\ref{fg_alpha_L}, so that $\alpha$ with different longitudinal fireball size $L$ can be compared. To be intuitive, we have replaced the entropy density on the $x$-axis by the temperature for $L=+\infty$ in the figure. In this calculation, we have taken $m_c=1.87$ GeV, and $m=3.1$ GeV, so that the threshold is $\epsilon=0.71$ GeV. The dissociation rate $\alpha$ increases with $L$ when $L$ is small up to about $L=0.8$ fm. Then it decreases a little at $L=1$ fm, and finally approaches to the curve with $L=+\infty$. For the curve $L=0.2$ fm, the energy of ground state of gluon is $\omega_0=\pi/L=3.1$ GeV$\gg \epsilon$, which leads to small cross section as indicated by Eq.~(\ref{eq_8}), and therefore the dissociation rate is far smaller than that in infinite space. At  $L=0.8$ fm, the corresponding $\omega_0=0.77$ GeV is comparable with $\epsilon$. In other words, the length that maximize the dissociation rate  can be estimated as $L_m={\pi}/\epsilon$ if the system is thermalized. This offers a time scale, far before which the gluon dissociation is negligible, even if the temperature of the fireball is high. Values of $L_m$ are listed in Tab.~\ref{tb_Lm}, where the binding energy is taken as that in vacuum.  Around $L_m$, almost all the gluons are above the threshold, but not too far from it, so that $\alpha$ is larger than that in infinite space in which case some gluons are with low energy, especially when the entropy density is relatively small. When $L$ exceeds $L_m$, part of gluons drops below the threshold, resulting in a smaller $\alpha$, as the curve of  $L=1$ fm shows. In this case, $\alpha$ is smaller at the high entropy density region than that in infinite space because of the valley structure in gluon spectrum, and it is larger at the low entropy density region because the energy of gluons are still remarkably larger than the thermal average in infinite space. (See Fig.~\ref{fg_gs})

\begin{table}[!hbt]
   \begin{tabular}{llllllllll}
	\hline
	quarkonium\rule{5mm}{0mm} & $J/\psi$&$\psi'$\rule{8mm}{0mm}&$\Upsilon(1S)$&$\Upsilon(2S)$&$\Upsilon(3S)$\\
	\hline
	$m$ (GeV)&$3.097$& $3.686$ &$9.460$ &$10.023$ &$10.355$\\
	$\epsilon$ (GeV)&$0.71$&$0.054$&$1.16$&$0.55$&$0.21$\\
	$L_m$ (fm)&$0.87$ & $11.6$& $0.53$ &$1.1$&$3.0$\\
	\hline
   \end{tabular}
   \caption{Threshold $\epsilon$ and the longitudinal size $L_m$ that approximately maximize the dissociation rate at fixed entropy density.}
   \label{tb_Lm}
\end{table}

With the dissociation rate, one finds the survival probability $R$ of a initially produced $J/\psi$ 
\begin{eqnarray}
	R&=&e^{-\int_{t_0}^{t_f}\alpha \dd t}.\label{eq_R}
\end{eqnarray}
In models without the finite volume effects, $t_0$ is usually taken as the formation time of the QGP to avoid the divergence of $\alpha$, and the suppression before the formation of the QGP is neglected. Therefore the survival probability depends on the parameter $t_0$. The results of Tab.~\ref{tb_Lm} implies that the time scale of gluon dissociation varies dramatically for different bound states. Since $\alpha$ vanishes at $t=0$, an alternative way to calculate the survival probability is to set $t_0=0$ in Eq.~(\ref{eq_R}) to include part of the dissociation before the formation of QGP and the parameter $t_0$ dependent is released. The following calculation adopted the later one. 

In order to understand the results of $R$, we work out three calculations for comparison: (I) taking $t_f$ as the end of the mixed phase (about 6.5 fm/$c$ in Fig.~\ref{fg_T}), when the QGP entirely vanish with $L$ dependent critical temperature $T_c$, (II) taking $t_f$ as the beginning of the mixed phase (about 2.0 fm/$c$ in Fig.~\ref{fg_T}) with $L$ dependent $T_c$, (III) taking $t_f$ as the end of the mixed phase (about 7.0 fm/$c$ in Fig.~\ref{fg_T}) with fixed $T_c=165$ MeV but $L$ dependent equation of state in QGP phase. The results are listed in Tab.~\ref{tb_R} with different initial entropy density, the entropy density is labeled by the temperature $T$ at time $t$ in the QGP phase without finite volume effects. By comparing (II) and (III), one finds that the shift of $T_c$ affect the survival probability of $J/\psi$ little. By comparing (I) and (II), one finds that the suppression is mainly due to the pure QGP phase, and the mixed phase plays a minor role under the conditions listed in the table. These results are observed with both $p_T=0$ GeV and $p_T=3$ GeV of $J/\psi$s, and with different entropy densities.
\begin{table}[!hbt]
   \begin{tabular}{l|lll|lll}
	   \hline
	   $p_T$ (GeV)& \multicolumn{3}{c|}{$0$} & \multicolumn{3}{c}{$3$}\\
	   calculation&(I) & (II)& (III)&(I)&(II)&(III)\\
	\hline
	$t=1.0$ fm $T=250$ MeV &0.482 & 0.412 & 0.416 &0.511 & 0.436 & 0.433\\
	$t=0.6$ fm $T=250$ MeV &0.693 & 0.622 & 0.627 &0.723 & 0.649 & 0.643\\
	$t=1.0$ fm $T=200$ MeV &0.743 & 0.688 & 0.686 &0.768 & 0.711 & 0.702\\
	\hline
   \end{tabular}
   \caption{Survival probabilities of $J/\psi$ at different calculations.}
   \label{tb_R}
\end{table}
\section{Summary}
Gluon dissociation of $J/\psi$ is studied in a box that is finite in the longitudinal direction and infinitely large in the transverse direction. The longitudinal boundary conditions change the spectrum of thermal partons, the equation of state of the medium and the critical temperature of the deconfinement phase transition. This modifies the values of the medium temperatures at the same entropy density, and therefore modifies the time evolution of the temperature under the assumption of entropy conservation.  With different values of the temperatures and the time evolution of the bulk medium, the $J/\psi$ dissociation rate and the survival probability are calculated.  The finite size effect reduces the $J/\psi$ suppression. When the medium size is around $L_m=\pi/\epsilon$, the corresponding gluon distribution located in this finite volume gives the strongest dissociation on  $J/\psi$ with its binding energy $\epsilon$.
\section*{Acknowledgment} This work is supported by the National Natural Science Foundation of China (NSFC) under Grant No. 12175165. 

\bibliographystyle{unsrt}
\bibliography{refs}
\end{document}